\documentclass[9pt]{article}
\usepackage{geometry}
\usepackage{amsmath,amssymb,hyperref}
\usepackage{graphicx,authblk} 
\usepackage{float}
\usepackage{fancyhdr}
\usepackage{array}
\usepackage[usenames,dvipsnames]{xcolor}

\usepackage{epstopdf}

\definecolor{cream}{RGB}{222,217,201}
\definecolor{linkColor}{rgb}{1,0,0}

\newcommand{\be}{\mathbf{e}}
\newcommand{\bn}{\mathbf{n}}

\newcommand{\zd}{0^\circ}
\newcommand{\sd}{60^\circ}
\newcommand{\otd}{120^\circ}

\newcommand{\tm}{\theta_{\mathrm{M}}}
\newcommand{\tl}{\theta_{\mathrm{L}}}
\usepackage{mathtools}

\begin{document}


\title{A Generalized Read-Shockley Model and Large Scale Simulations for the Energy and Structure of Graphene Grain Boundaries}
\author[1]{Ashivni Shekhawat}
\author[2]{Colin Ophus}
\author[3,4]{Robert O.~Ritchie}
\affil[2]{National Center for Electron Microscopy, Molecular Foundry, Lawrence Berkeley National Laboratory, Berkeley}
\affil[1]{Miller Institute for Basic Research in Science, University of California Berkeley}
\affil[3]{Department of Materials Science and Engineering, University of California Berkeley}
\affil[4]{Materials Sciences Division, Lawrence Berkeley National Laboratory, Berkeley}


\maketitle
\abstract{The grain boundary (GB) energy is a quantity of fundamental importance for understanding several key properties of graphene. Here we present a comprehensive theoretical and numerical study of the entire space of symmetric and asymmetric graphene GBs. We have simulated over 79,000 graphene GBs to explore the configuration space of GBs in graphene. We use a generalized Read-Shockley theory and the Frank-Bilby relation to develop analytical expressions for the GB energy as a function of the misorientation angle and the line angle, and elucidate the salient structural features of the low energy GB configurations. } \\

\section{Introduction}
Graphene -- a two dimensional allotrope of carbon with excellent mechanical and electronic properties -- has attracted much attention since it was first produced by direct exfoliation from graphite more than a decade ago~\cite{novoselov2009,geim2007,novoselov2005,balandin2008,li2009large,bae2010roll,liu2012biological,bonaccorso2010graphene,sun2008,ren2014global}. Exfoliated graphene is largely monocrystalline; however, exfoliation is not a scalable production technique. Chemical vapor deposition (CVD) is the most widely used scalable production technique, and is being used to produce more than 300,000 m$^2$ of graphene annually~\cite{li2009large,bae2010roll,ren2014global}. Graphene produced from CVD is polycrystalline~\cite{yazyev2014}, and thus it contains intrinsic line defects in the form of grain boundaries (GBs) that have been studied observed experimentally~\cite{huang2011grains,an2011domain,kim2011grain,kurasch2012atom,rasool2014conserved,tison2014grain}. Such GBs have a profound effect on the properties of the polycrystalline materials; for instance, a high GB energy can promote grain growth, or a low GB strength can lead to brittle intergranular fracture.  It has also been noted that while some graphene GBs offer minimal resistance to electron transport, other GBs can be highly insulating~\cite{yazyev2010electronic}. Several recent studies have demonstrated that the thermal conductivity of graphene is strongly dependent on the GB structure~\cite{bagri2011thermal,serov2013effect,grosse2014direct,yasaei2015bimodal}.  The mechanical strength of graphene is  influenced by GBs~\cite{wei2012nature,yi2013theoretical,zhang2012,grantab2010}, which can have a profound influence on sustainability application such as sea water purification by reverse osmosis.  Additionally, the GB structure in polycrystalline single-layer graphene strongly modifies electronic transport \cite{koepke2013atomic,Ma2014226802,ma2015structural,ADMA201401389,kendal2013}.

Thus, it is clear that in order to understand the properties of polycrystalline graphene, it is necessary to characterize graphene GBs. Perhaps the most important property of a GB is its excess energy (per unit length). The excess GB energy, or simply the GB energy, has direct influence on the grain morphology~\cite{Mishin20101117,Rollett19891227}, and thus influences all grain morphology dependent properties including strength and transport. It is not feasible to measure the GB energy directly in an experiment. Instead, GB energy in crystalline materials is typically inferred from the equilibrium structure of triple junctions (interface of three GBs)~\cite{herring1951some,hasson1972theoretical,adams1999extracting}. Such equilibrium junctions satisfy the Herring equations~\cite{herring1951some}, which can be used to deduce the GB energy if a statistically large amount of experimental data is available~\cite{adams1999extracting}. No such experimental study has been performed for graphene. Indeed, we (with co-authors) published the first statistically large dataset of high-resolution transmission electron microscopy (HRTEM) observation of graphene GBs only recently~\cite{ophus2015large}. Neither us, nor any other group has reported a statistically relevant number of experimental observations of triple junctions in graphene. On the other hand, computer simulations can be used to directly measure the GB energy without resorting to the indirect method of triple junctions~\cite{Mishin20101117}. Although there have been several numerical studies of graphene GBs, most of these studies have focused on a few special configurations or symmetric GBs, and have not explored the entire configuration space of graphene GBs~\cite{yazyev2010topological,Liu20112306,carlsson2011theory,liu2010cones,malola2010structural,zhang2013structures,Romanov2015223,ADFM:ADFM201403024,C5NR04960A,Han2014250,C3NR06823D}. Here we present a comprehensive numerical and theoretical study of the energy of a very large set of graphene GBs. We have used molecular dynamics (MD) simulations to measure the energy of about 79,000 grain boundary configurations, corresponding to 4122 unique $(\tm,\tl)$ points, and spanning the entire space of all possible graphene grain boundaries. 

Traditionally, the energy of low angle grain boundaries in bulk materials is understood in terms of Read and Shockley's theory~\cite{read1950dislocation,sutton1995interfaces,GuiJin1986}. This theory was developed for bulk materials, and is not directly applicable to two dimensional membranes. This is due to the fact that the individual dislocation cores in a two dimensional membrane can buckle out of plane - a mode of relaxation that is not available in their three dimensional counterparts. By buckling out of plane, the dislocation can trade in-plane strain for out of plane bending, thereby lowering its energy considerably~\cite{nelson,yazyev2010topological,chen2011continuum}. We use a generalized Read-Shockley theory for two dimensional membranes, and combine it with the Frank-Bilby~\cite{frank1953martensite,bilby1955continuous,hull1984introduction,sutton1995interfaces,hirth2013interface} equation to elucidate the structure of the energy function of all possible graphene GBs. Further, we develop a theoretical understanding of the salient structural features of graphene GBs. Our results should be applicable to a large class of 2D materials, and will lead to a better understanding of fundamental processes such as grain growth, transport, and strength in these materials.
\section{Modeling}
\subsection{Configuration Space of Graphene GBs}
\begin{figure*}
\centering
\includegraphics[width=0.9\linewidth]{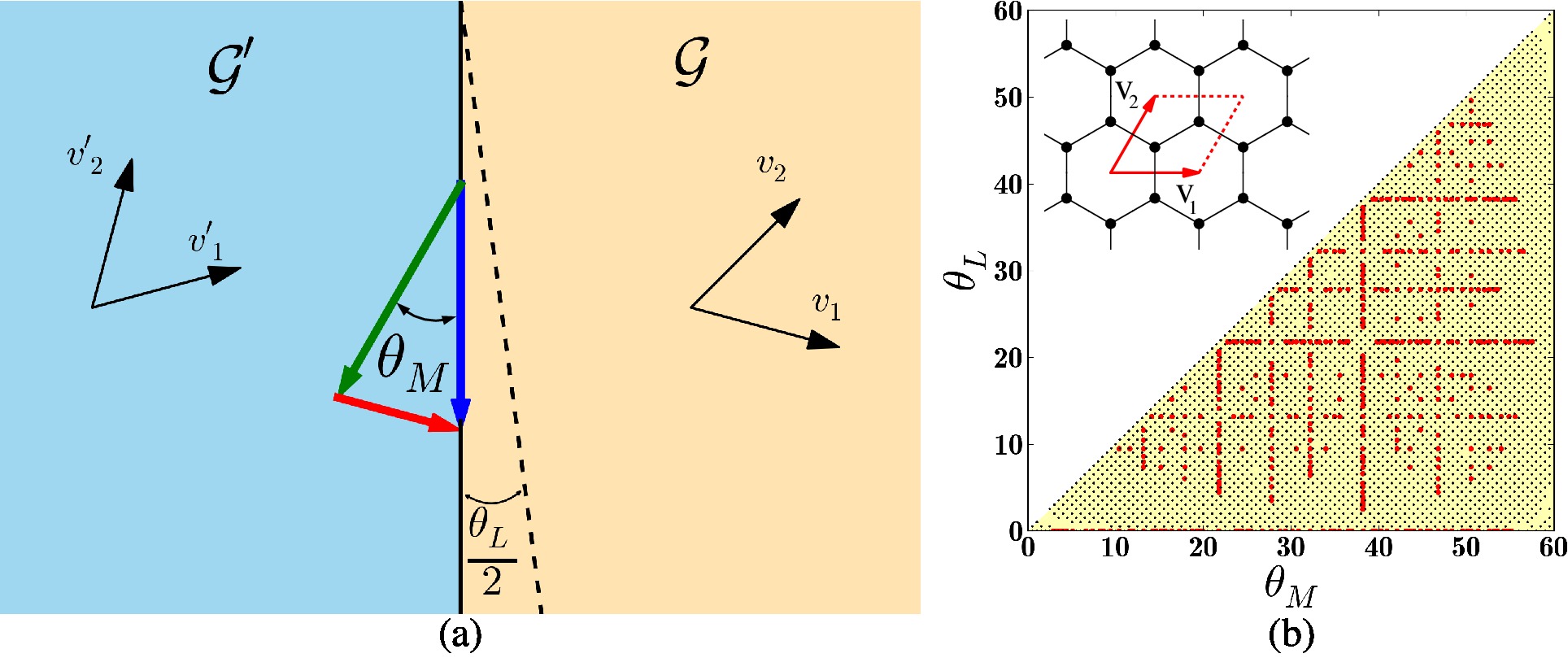}
\caption{(Color Online) (a) Two grains $\mathcal{G},\ \mathcal{G}'$ and the GB interface (black line) between them. The half line angle, $\tl/2$, measures the deviation of the interface from the symmetric GB (dashed black line). The Burger's vector needed to complete the interfacial Burgers circuit is shown by the red arrow. (b) The space of unique GBs in graphene. The colored (yellow) triangular region contains all unique $(\tm,\ \tl)$ pairs (up to symmetry). The red circles and the black dots show all commensurate, and approximately commensurate GBs, respectively, with a repeat distance smaller than 2000\ \AA\ that were simulated in this study. The inset shows the lattice vectors and the primitive unit cell of graphene.}
\label{fig:figure1}
\end{figure*}
Figure~\ref{fig:figure1}a shows a general GB at the interface of two grain $\mathcal{G}$ and $\mathcal{G}'$, with lattice vectors $(v_1,\ v_2)$ and $(v'_1, v'_2)$, respectively (the inset in Figure~\ref{fig:figure1}b shows the lattice vectors and the primitive unit cell of graphene). GBs in graphene are characterized by two angles, namely, the misorientation angle $\tm$, and the line angle $\tl$. The misorientation angle characterizes the relative rotation of the two grains, i.e., $v'_i = R_{\tm}v_i$, where $R_{\tm}$ represents a positive rotation by $\tm$, whereas the line angle characterizes the deviation of the GB from the line of symmetry between the two grains. For any given $(\tm,\ \tl)$ pair there is a third degree of freedom given by the relative sliding of the grains along the GB, however, we choose the sliding that gives the lowest GB energy, thereby effectively eliminating this degree of freedom. 
Due to the symmetries of the graphene lattice, the space of unique $(\tm,\ \tl)$ pairs is reduced to a triangular area~\cite{ophus2015large}, as shown in Figure~\ref{fig:figure1}b.
Commensurate GBs (CSL) exist at certain special values of $(\tm,\ \tl)$, while an approximately commensurate GB can be constructed at any $\tm,\ \tl$~\cite{ophus2015large,sethna2008,sutton1995interfaces}. We simulate all commensurate GBs with a repeat length less than 2000\ \AA~\cite{ophus2015large}. Further, we grid the $(\tm,\ \tl)$ space in steps of 0.5$^\circ$ with approximately commensurate GBs. For each unique $(\tm, \tl)$ pair, several simulations have to be performed to explore the relative sliding between the two grains; in all we have simulated and evaluated the energy of over 79,000 GB configurations corresponding to 4122 unique $(\tm,\ \tl)$ pairs. The details of the GB configurations and structures used in this study can be found in Ref.~\cite{ophus2015large}. The excess energy per-unit-length of a GB is calculated as
$\gamma(\tm,\tl) = (E_{\mathrm{total}} - n_{\mathrm{atoms}}E_{\mathrm{bulk}})/l_\mathrm{GB}$,
where $E_{\mathrm{total}}$ is the net energy of the configuration, $n_{\mathrm{atoms}}$ is the number of atoms in the configuration, $E_\mathrm{bulk}$ is the energy per-atom  in the reference crystal (= -7.81 eV for the AIREBO potential), and $l_\mathrm{GB}$ is the length of the GB. We use the AIREBO potential~\cite{stuart2000,stuart2002} as implemented in the LAMMPS code~\cite{plimpton1995}, and all our GB configurations are thoroughly relaxed, and allow for out of plane deformations (see the Methods Section for details). The GB structures used in this study are available online~\cite{onlineRepo}.

\subsection{Dislocation Model For GBs}
Before discussing the GB structures and energy in detail, we present a dislocation based model for the GBs. This model will be used to elucidate the structure of the GBs, and to derive functional forms for the GB energy. The Frank-Bilby equation can be used to calculate the interfacial Burger's vector (per-unit-length) of the geometrically necessary dislocations for a GB with misorientation $\tm^0$ and line angle  $\tl^0$. This is the Burger's vector required to close the Burger's circuit shown in Figure~\ref{fig:figure1}a (the red arrow), and is given by $\bn^0 = 2\sin(\tm^{0'}/2)\left( \cos(\tl^{0}/2)\be_1 - \sin(\tl^0/2)\be_2\right)$, where, $ \tm^{0'} = \tm^0 $ for $0< \tm^0 \leq 3\zd$, while $ \tm^{0'} = 6\zd - \tm^0$ for $3\zd< \tm^0 \leq 6\zd$, and, $\be_{1,2}$ are the unit vector parallel and perpendicular to $v_1$, respectively (Supporting Information Section S1). Let the energy of this GB be $\gamma^0 = \gamma(\tm^0,\ \tl^0)$. Now consider a GB near this configuration, with the perturbed misorientation and line angle given by $(\tm^0 + \delta \tm,\ \tl^0 + \delta \tl)$. The perturbation in the density of interfacial Burger's vector is given by $\delta \bn = (\partial \bn^0/\partial \tm^0)\delta \tm + (\partial \bn^0/\partial \tl^0)\delta \tl$. We assume that this change in Burger's vector density is accommodated by well separated (1,0) dislocations introduced along the boundary. There are three independent (1,0) dislocations in the reference crystal, with Burgers vectors in the directions $R_{\zd, \sd, \otd}\be_1$; thus $\delta \bn = \delta n_1 \be_1 + \delta n_2 R_{\sd} \be_1 + \delta n_3 R_{\otd} \be_1$, giving
\begin{multline}
\delta n_1 \be_1 + \delta n_2 R_{\sd} \be_1 + \delta n_3 R_{\otd} \be_1 = \\
  (\partial \bn^0/\partial \tm^0)\delta \tm + (\partial \bn^0/\partial \tl^0)\delta \tl,
\label{eq:Cons}
\end{multline}
where $\delta n_2$ is the perturbation in the density of Burger's vector due to dislocations in the $R_{\sd} \be_1$ direction, etc. 
The above vector equation provides two constraints on the perturbation of the density of three independent (1,0) dislocations, thus leaving the system indeterminate. We obtain one more condition by writing a perturbed GB energy and minimizing it with the above constraints. Since the perturbation is small, the new dislocations introduced into the GB are well separated and do not interact. In the traditional Read-Shockley theory, the energy of an isolated dislocation has a divergent logarithmic term~\cite{read1950dislocation,hull1984introduction}. This term is due to the fact that, in a bulk material, the long range strain field of an isolated dislocation decays with distance as $1/r$. However, it is known that in a two dimensional membrane the bending stiffness is small, and it is energetically favorable to trade long range strain for out of plane deformation, thereby removing the logarithmic term from the energy of the isolated dislocation core~\cite{nelson,chen2011continuum}. It can be shown that for two dimensional membranes, each isolated dislocation costs a finite (constant for a given GB) amount of energy~\cite{nelson,chen2011continuum}. Thus, we can write the following minimization problem for the perturbation $\delta \bn$
\begin{multline}
\mathrm{Min.}\ \gamma(\tm^0 + \delta \tm, \tl^0 + \delta \tl) = \gamma(\tm^0, \tl^0) +\\
+ \frac{Gb}{4\pi(1-\mu)} \Sigma c_i | \delta n_i |,\quad \mathrm{subject\ to\ constraints\ \ref{eq:Cons},}
\label{eq:Hamiltonian}
\end{multline}
where $G$ is the shear modulus, $\mu$ is the Poisson's ratio, $b$ is the Burger's vector, and $c_i$'s are dimensionless constants for a given configuration $(\tm^0,\tl^0)$ representing the energy required to embed the (1,0) dislocations into the GB. The validity of such a perturbational form for the GB energy has been tested numerically and experimentally~\cite{gjostein1959absolute,GuiJin1986}. If $c_i$'s are known, then the minimization of the perturbed energy in Equation~\ref{eq:Hamiltonian} with respect to the dislocation density $\delta n_i$ can be performed analytically. However, the coefficients $c_i$'s are unknown functions of $(\tm^0,\tl^0)$, and thus a close form solution to the minimization problem is infeasible. Yet, we will show that considerable insight can be gained from this formulation. 
\begin{figure*}[tbp]
\centering
\includegraphics[width=0.8\linewidth]{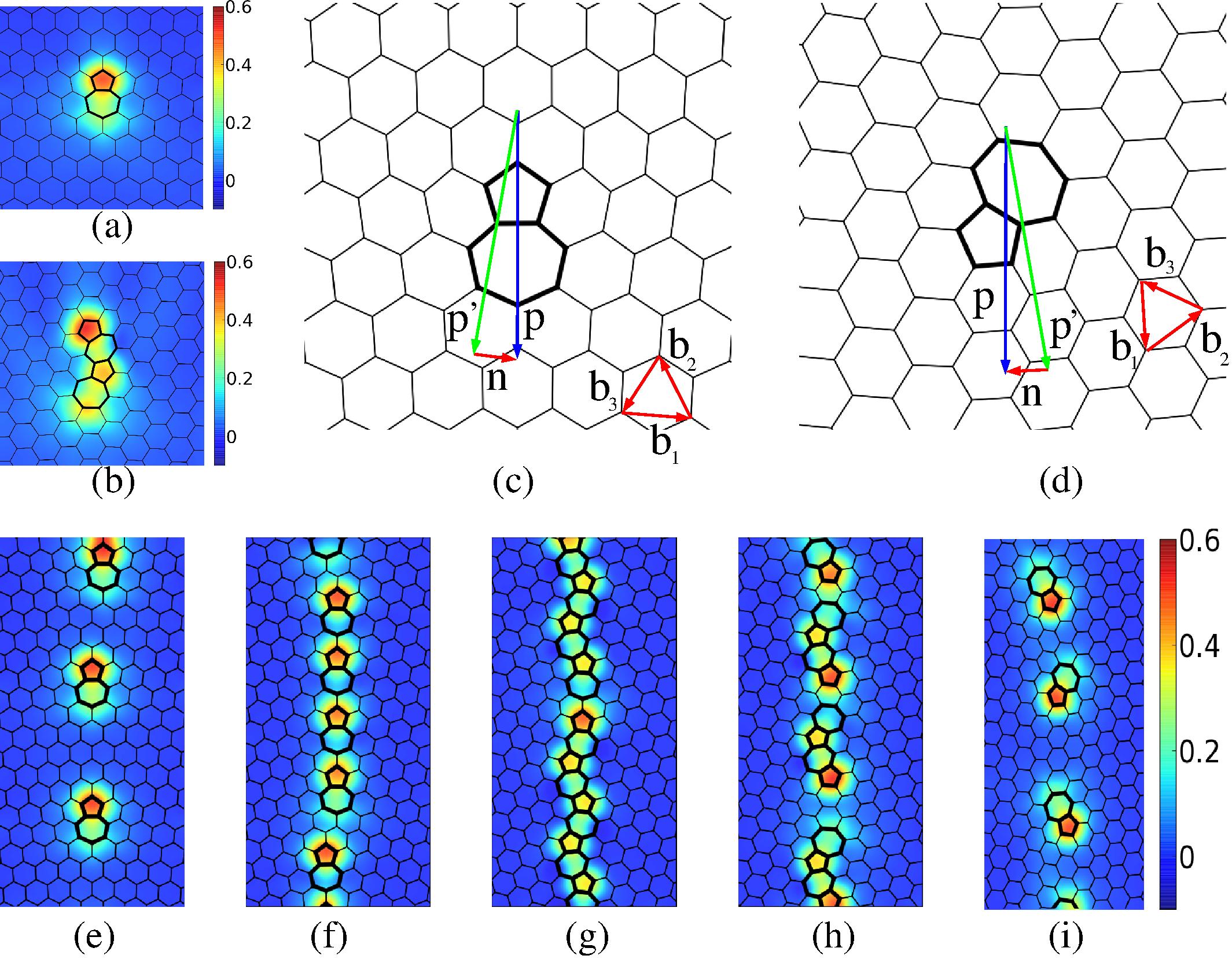}
\caption{(Color Online) The structure and crystallography of graphene GBs. (a), (b) show the dislocation core of isolated (1,0) and (1,0)+(0,1) dislocations, respectively. The color indicates the excess energy per atom in units of eV. (c), (d) show isolated dislocations at a low angle ($\tm = 1\zd$) and high angle ($\tm = 5\zd$) symmetric GB. The shortest interfacial Burger's vector is indicated by $\bn$. (e)-(i) show sections of the lowest energy GBs at $\tm = 1\zd, 2\zd, 3\zd, 4\zd, 5\zd$; color scheme same as (a), (b). }
\label{fig:figure2}
\end{figure*}

\section{Results and Discussions}
\subsection{Structure And Energy of Symmetric GBs}
Graphene GBs are known to be composed of rings of five and seven carbon atoms (apart from the usual hexagonal rings)~\cite{yazyev2010topological,huang2011grains,ophus2015large}. These pentagon-heptagon pairs form the cores of the dislocations with the shortest Burger's vector. Figure~\ref{fig:figure2}a,b show the dislocation core with Burger's vector (1,0) and (1,0)+(0,1)~\cite{yazyev2010topological}. Figure~\ref{fig:figure2}e-i show segments of symmetric GBs with $\tm = 1\zd, 2\zd, 3\zd, 4\zd$, and $5\zd$. It can be seen that the GBs are composed of (1,0) or (1,0)+(0,1) dislocation cores. The GBs with low misorientation ($\tm = 1\zd, 2\zd$) are composed of (1,0) dislocations whose Burger's vectors are aligned in a single direction. Analogously, the GBs with large misorientation ($\tm = 5\zd$) are composed of (1,0) dislocations whose Burger's vectors alternate by $\sd$ in their orientation. As mentioned previously, the graphene lattice has three independent (1,0) dislocations whose Burger's vectors are rotated by $\sd$ with respect to each other (the system $b_{1,2,3}$ in Figure~\ref{fig:figure2}c,d, for example). Thus, while in principle a general symmetric GB can have (1,0) dislocations with three different orientations, the energy minimizing configurations of low angle symmetric GBs have their (1,0) dislocations aligned along just one direction; furthermore, the high angle symmetric GBs have dislocations aligned with two of the three possible directions. These structural features can be explained by considering the symmetric low and high angle GBs as perturbations about the pristine crystal (obtained at $\tm = \zd, \sd$), and solving the minimization problem given by Equation~\ref{eq:Hamiltonian} (details in Supporting Information Section S2). The essential insight is that for low angle GBs the perturbed interfacial Burger's vector is almost parallel to the lattice vector $v_1$ (or equivalently to the lattice Burger's vector $b_1$, see Figure~\ref{fig:figure2}c). Thus, the energy minimizing configuration results when all the (1,0) cores are aligned with $v_1$ (equivalently, $b_1$). Aligning the core with $R_{\sd}\be_1$ instead, for example, would need twice the number of  dislocations, and hence would cost twice the amount of energy, and thus would be suboptimal. Similarly as it can be seen that for high angle GBs the perturbed interfacial Burger's vector is almost perpendicular to $v_1$ (equivalently $b_1$, Figure~\ref{fig:figure2}d),  it is energetically not beneficial to have a dislocation Burger's vector aligned with $v_1$ (or $b_1$) . Hence, the high angle GBs have dislocations with Burger's vectors aligned with $b_2$ and $b_3$ only. 
\par
\begin{figure*}[tbp]
\centering
\includegraphics[width=0.9\linewidth]{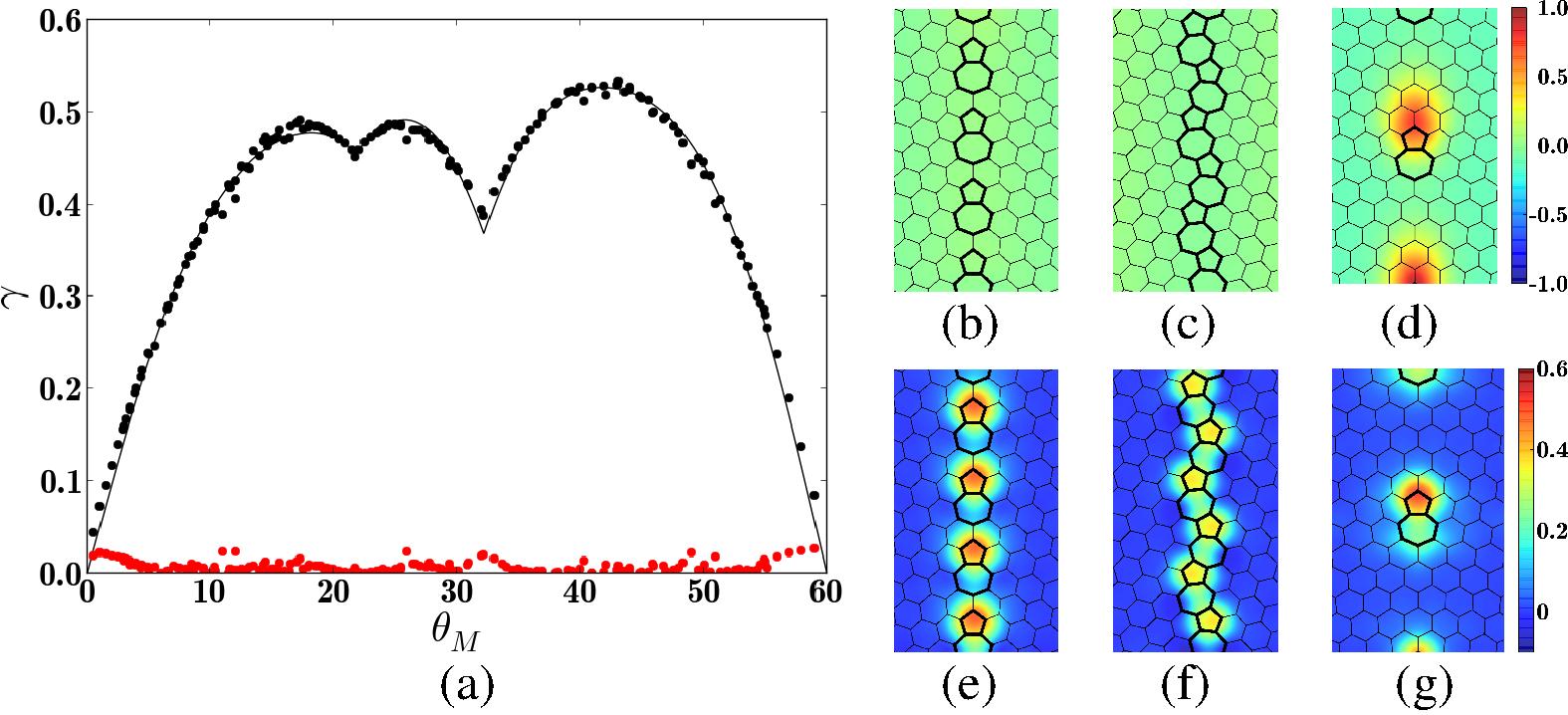}
\caption{(Color Online) (a) The energy of all simulated symmetric GBs in units of eV/\AA. The filled black circles show the simulation data, while the solid line is a fit to Equation~\ref{eq:symGBEnergy}. The filled red circles show the magnitude of the fitting error, which is smaller than 0.06 eV/\AA\ everywhere. (b), (c) show the $\Sigma_{7,\ 13}$ GBs, and (d) shows a symmetric GB with $\tm = 1\zd$, colored by the net out of plane displacement in unit of\ \AA. (e)-(g) show the same GBs colored by excess energy per-atom in units of eV.}
\label{fig:figure3}
\end{figure*}

Figure~\ref{fig:figure3}a shows the numerically measured energy function for symmetric GBs, i.e., $\gamma(\tm, 0)$. It is well known that the GB energy has cusps at special high symmetry (low $\Sigma$ CSL, where CSL stands for the `Coincident Site Lattice', and $\Sigma$ denotes the ratio of the volume of the unit cell of the CSL to that of the regular lattice, see Ref.~\cite{sutton1995interfaces} for a detailed discussion of CSL) boundaries~\cite{Wolf1990781,duffy,TSUREKAWA1994341,sutton1995interfaces,bulatov2014grain}, and these can be seen clearly in Figures~\ref{fig:figure3}a,~\ref{fig:figure4}a. There are two prominent cusps for graphene~\cite{Liu20112306}: first at the $\Sigma_7 (\tm = 21.78^\circ,\ \tl = \zd)$ GB, and the second at $\Sigma_{13}(\tm = 32.2^\circ,\ \tl=\zd)$ GB. Apart from these, there are the obvious families of cusp singularities at $\tm = \zd,\ \sd$. The $\Sigma_{7,\ 13}$ GBs are strong local energy minima of the GB energy. These minima arise due to favorable interactions between the dislocation cores. For intermediate values of misorentation $(15^\circ \lesssim \tm \lesssim 45^\circ)$, the density of required dislocations is high, and the individual cores cannot be well separated. We note that for isolated (1,0) as well as (1,0)+(0,1) cores, there is compression at the tip of the leading pentagonal ring and dilation at the tail of the trailing heptagonal ring (seen by the relative shortening and stretching of the bonds, most clearly visible in Figure~\ref{fig:figure2}a,b)~\cite{wei2012nature}. This local straining leads to significant out of plane buckling near the dislocation, as seen in Figure~\ref{fig:figure3}d~\cite{yazyev2010topological}. However, as the dislocation density increases with increasing $\tm$, and two (1,0) or (1,0)+(0,1) approach each other, their strain fields cancel, and there is a reduction in the elastic energy of the system. This cancellation of strain fields can be inferred from Figures~\ref{fig:figure3}b,c, where it can be seen that the $\Sigma_{7,\ 13}$ GBs have almost no out of plane buckling, because the strain fields cancel out very effectively in these GBs with tightly arranged dislocations. On the other hand, at higher $\tm$ (note that the dislocation density peaks at $\tm = 3\zd$), the increased density of the dislocations leads to higher energy per-unit-length. Thus, there is a competition between the energy increase due to higher dislocation density, and energy decrease due to dislocation interaction. It can be seen that initially the GB energy increases with $\tm$, thus the energy increase dominates over the energy reduction. However, the reduction becomes significant, and the net energy starts to decrease at about $\tm = 18^\circ$. This reduction in energy reaches a first optimum for the $\Sigma_7 (\tm = 21.78^\circ)$ CSL GB (Figure~\ref{fig:figure3}b,e) where all (1,0) dislocation pairs are aligned, and there is a separation of exactly 1 carbon-carbon bond between them. At this optimal configuration there is significant reduction in the elastic energy, resulting in the first cusp in the GB energy (Figure~\ref{fig:figure3}a). Increasing the misorientation $\tm$ further initially leads to an increase in the GB energy. This is due to the fact that a higher dislocation density pushes dislocations closer; however, geometrically it is still favorable to have all dislocations aligned in the same direction. Thus, creating a (1,0)+(0,1) pair incurs an energy penalty. However, with further increase in $\tm$, it becomes progressively more favorable for the individual dislocations to stagger and merge to form (1,0)+(0,1) cores. This process leads to a reduction in energy starting at about $\tm = 24^\circ$. An optimal configuration is reached at the $\Sigma_{13} (\tm = 32.2^\circ)$ CSL GB where all (1,0)+(0,1) line up perfectly (Figure~\ref{fig:figure3}c,f), and leads to a large reduction in the elastic energy, resulting in the second, deeper cusp in the GB energy. On increasing $\tm$ further, the net dislocation density decreases and the (1,0)+(0,1) cores separate, ultimately resulting in the behavior for large misorientations discussed previously. 

Having understood the most salient features of the GB structure, we now turn our attention to the GB energy. A simple analysis of Equation~\ref{eq:Hamiltonian} shows that for 2D materials the GB energy function has a absolute value ($|\cdot|$) type singularity at the cusps (Supporting Information section S2). Thus, we propose the following functional form for the energy of the symmetric GBs:
\begin{multline}
\gamma_{\mathrm{sym}}(\tm) = \frac{Gb}{4\pi(1-\mu)}\big| \sin 3\tm\big| \Big( \Sigma_{i=2}^n p_i\cos 3i\tm +\\
+\Sigma_{i=1}^{n_c} a_i\big|\cos 3\tm - \cos 3\tm^{c_i}\big|\Big),
\label{eq:symGBEnergy}
\end{multline}
where $p_i,\ a_i$ are dimensionless fitting parameters. The overall factor of $|\sin 3\tm|$ gives the correct asymptotic form at the cusps at $\tm = \zd,\ \sd$. The first term ($p_i$'s) fits the smooth variation in the energy function. The second term ($a_i$'s) fits the cusps at angles $\tm^{c_i}$. Although any desired number of cusps can be included, we include the two prominent cusps at the $\Sigma_{7,\ 13}$ CSL GBs, thus $n_c=2$, and $\tm^{c_{1,2}} = 21.78^\circ,\ 32.2^\circ$. Note that this form satisfies all symmetry requirements, namely a period of $\otd$, and even reflection symmetries about $\tm = \zd$ (i.e., $\gamma_{\mathrm{sym}}(\tm)$ $= \gamma_{\mathrm{sym}}(-\tm)$ $= \gamma_{\mathrm{sym}}(\otd + \tm)$), and has the correct asymptotic form near the cusp singularities. We do not include the $n=0,1$ terms because the corresponding harmonics are included in the expression for the cusps. The solid line in Figure~\ref{fig:figure3}a shows a fit of Equation~\ref{eq:symGBEnergy} to the simulation data with $n = 4$, giving a total of just 5 fitting parameters. The values of these parameters at the best fit are $p_2 = -3.70\times10^{-2},\ p_3 = 6.18\times10^{-3},\ p_4 = 1.99\times10^{-2},\ a_1 = 8.91\times10^{-2},\ a_2 = 2.00\times10^{-1}$, while $G,\ \mu$ are measured to be $325.68$ GPa, and $0.318$, respectively, from separate MD simulations (Methods Section). The maximum absolute error for the fit is 0.026 eV/\AA. It is clear that the theory provides an excellent fit to the data with a minimal number of fitting parameters. Including the higher  harmonics (bigger $n$) does not result in a significant improvement in the results (Supporting Information Figure S1). We find that fitting a Fourier series without including the correct asymptotic form of the cusps results in very poor performance; fits with as many as 50 free Fourier components are needed for an accuracy similar to our fit with 5 parameters (Supporting Information Figure S2). Finally, we note that our functional form is reminiscent of the form used by Sethna and Coffmann~\cite{sethna2008}, however, their form, while more pedagogical, had several redundant parameters (a total of 16 parameters, as opposed to our 5), and thus did not provide a minimal description of the GB energy. 
\subsection{Energy of  Asymmetric GBs}
\begin{figure*}[tbp]
\centering
\includegraphics[width=0.9\linewidth]{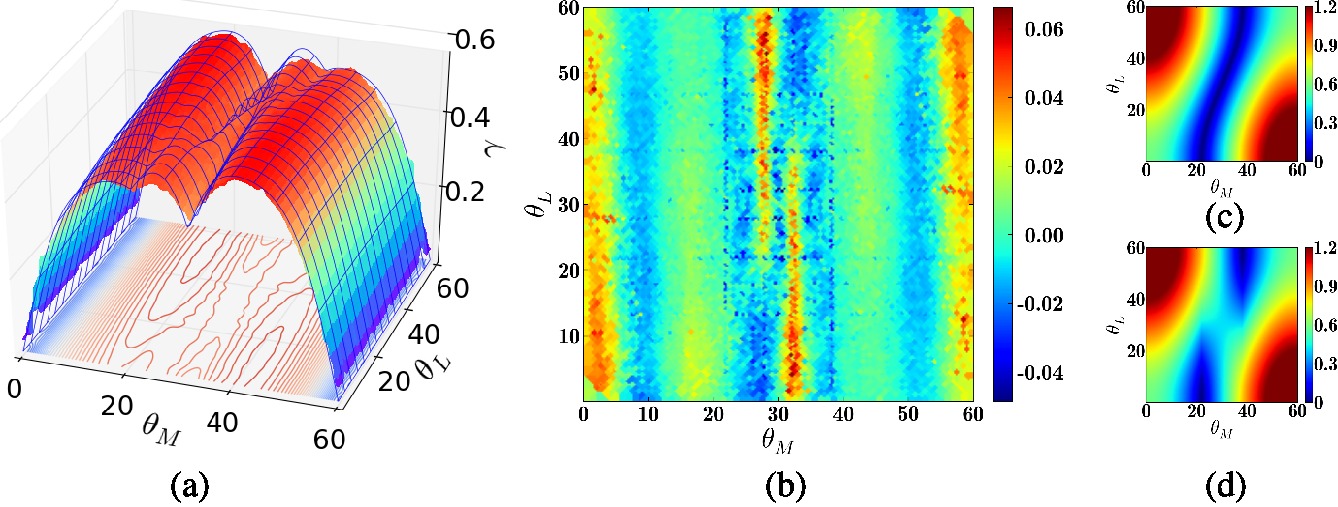}
\caption{(Color Online) (a) Energy of all simulated symmetric and asymmetric GBs in units of eV/\AA. The surface shows data from simulation, while the grid is a fit to Equation~\ref{eq:FullGBEnergy}. (b) The error in fit shown in units of eV/\AA. (c), (d) The basis functions, $|h_1(\tm,\tl,\tm^{c}) + h_2(\tm,\tl,\tm^{c})|$, $|h_1(\tm,\tl,\tm^{c})| +$ $ |h_2(\tm,\tl,\tm^{c})|$ used to fit the cusp singularity at $\tm^c = 21.78^\circ$ in Equation~\ref{eq:FullGBEnergy}.}
\label{fig:figure4}
\end{figure*}

Figure~\ref{fig:figure4}a shows the numerically measured energy function $\gamma(\tm, \tl)$. It can be seen that the variation of $\gamma(\tm, \tl)$ in the $\tl$ direction is significantly smaller than that in the $\tm$ direction. This is a direct consequence of the fact that the magnitude of the interfacial Burger's vector, $\bn$, is given by $2\sin(\tm/2)$ and is independent of $\tl$. Since the energy is largely a function of the magnitude of the interfacial Burger's vector, it follows that the energy variation in the $\tl$ direction is smaller. However, the cusps in the symmetric energy function $\gamma_{\mathrm{sym}}(\tm, \zd)$ become ridges in the full energy function $\gamma(\tm, \tl)$. Thus, even though the energy function  varies slowly in the $\tl$ direction, its structure is made interesting by the presence of these ridges, particularly due to the symmetry requirement $\gamma(\tm, \tl) = \gamma(\sd - \tm, \sd-\tl)$ which makes the ridges turn (or vanish). 
The numerical data suggest that there are two kinds of ridges: one that join cusps at $(\tm^0,0)$ to its periodic counterpart at $(\sd - \tm^0, \sd)$, and another that continues almost straight in the $\tl$ direction without bending. Based on an analysis of Equation~\ref{eq:Hamiltonian}, we propose the following form for the general GB energy to captures these features (see Supporting Information Section S2 for details)
\begin{multline}
\gamma(\tm, \tl) = \frac{Gb}{4\pi(1-\mu)} \big| \sin 3\tm\big | \Big( \Sigma_{i=0}^n\Sigma_{j=0}^m p_{ij}I_{ij}\cos 3i\tm\cos 3j\tl \\
+ \Sigma_{i=1}^{k} \Big( a_i^t |h_1(\tm,\tl,\tm^{c_i}) + h_2(\tm,\tl,\tm^{c_i})| \\
+ a_i^s\left( |h_1(\tm,\tl,\tm^{c_i})| + |h_2(\tm,\tl,\tm^{c_i})|\right) \Big)\Big),\label{eq:FullGBEnergy}
\end{multline}
where $p_{ij},\ a_i^t,\ a_i^s$ are fitting parameters, $I_{ij} \equiv (1 + (-1)^{i+j})/2$ is an indicator function that is 1 if both $i,\ j$ are even or odd, and zero otherwise, $h_1(\tm,\tl,\tm^{c_i}) \equiv (\cos3\tm - \cos3\tm^{c_i})\cos^2 1.5\tl$, and $h_2(\tm,\tl,\tm^{c_i}) \equiv (\cos3\tm + \cos3\tm^{c_i})\sin^2 1.5\tl$. The indicator function is needed to make sure that the symmetry requirement $\gamma(\tm, \tl) = \gamma(\tm+\sd, \tl+\sd)$ is satisfied. Note that the functional form satisfies all other symmetry requirements as well (overall period of $\otd$ in $\tm,\ \tl$, even mirrors at $\tm, \tl = \zd$ and $\sd$, i.e., $\gamma(\tm,\tl)$ $= \gamma(-\tm,\tl)$ $= \gamma(\tm,-\tl)$ $= \gamma(\otd + \tm,\tl)$ $= \gamma(\tm,\otd+\tl)$ $= \gamma(\sd - \tm,\sd - \tl)$). We also set $p_{00} = p_{10} = p_{01} = p_{11} = 0$ because the constant term is not needed, and the other harmonics are contained in $h_i's$. The $a^t_i$ terms model ridges that turn, while $a^s_i$ terms model the ridges that remain straight. Taking $n,\ m = 4,\ 4$ and $k=2$ as before, gives a fit with 15 free parameters, which is presented in Figure~\ref{fig:figure4}a. The values of the parameters for the best fit can be found in the Supporting Information Section S3. The maximum absolute error of fitting is 0.07 eV/\AA, indicating a good quality fit. Figure~\ref{fig:figure4}b shows the fitting error. As before, adding further harmonics does not improve the quality of the fit significantly.  

\section{Conclusion}
\begin{figure*}[htbp]
\includegraphics[width=0.9\linewidth]{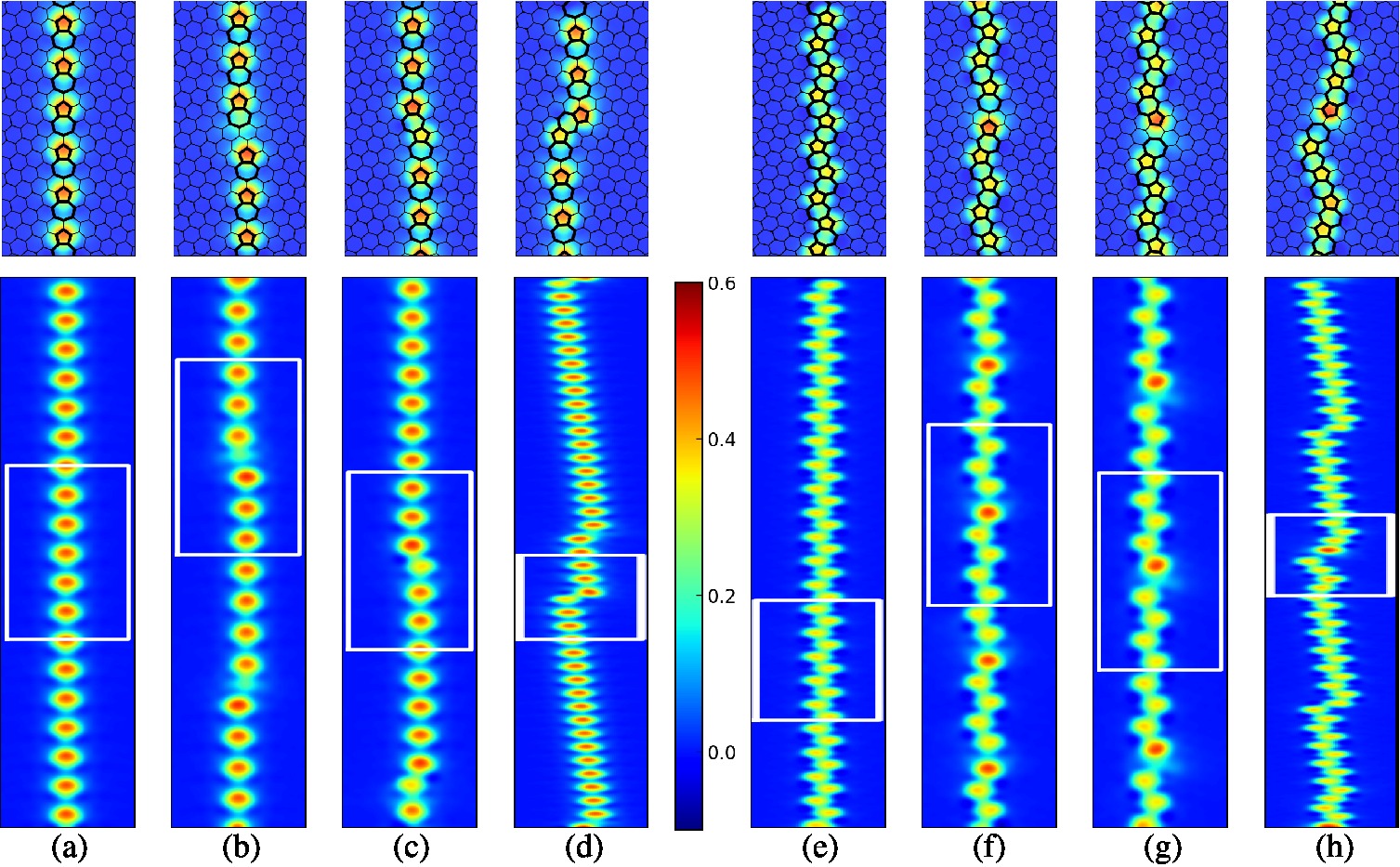}
\caption{(Color Online) Perturbations about the symmetric $\Sigma_{7,\ 13}$ GBs. The top panel shows a zoom-in of the atoms in the white box in the panel directly below. Color shows the excess energy per-atom in units of eV. (a) Several repeats of the coincident site lattice (CSL) unit cell of the $\Sigma_7$ GB. (b)-(d) Perturbations about the $\Sigma_7$ GB with $(\delta \tm,\ \delta \tl)$ $=(-0.78^\circ, \zd)$, $=(0.72^\circ, \zd)$, and $=(\zd, 4.84^\circ)$, respectively. (e) Several repeats of the CSL unit cell of the $\Sigma_{13}$ GB. (f)-(h) Perturbations about the $\Sigma_7$ GB with $(\delta \tm,\ \delta \tl)$ $=(-2.2^\circ, \zd)$, $=(1.8^\circ, \zd)$, and $=(\zd, 6.4^\circ)$, respectively.}
\label{fig:figure5}
\end{figure*}
We find that the Read-Shockley type dislocation model provides an accurate description of the structure and energy of graphene GBs. The functional forms for energy derived on the basis of this formulation are numerically efficient, containing just 5 fitting parameters for the symmetric GB energy, and 15 fitting parameters for the entire GB space. The absolute error in our fits is smaller than 0.07 eV/\AA\ everywhere. We find that main source of this error is the limited size of our simulation cells (due to computational limitations). It can be seen in Figure~\ref{fig:figure4}b that the largest error is concentrated in narrow bands around $\tm = \zd,\ \sd,\ 32.2^\circ$. These are high symmetry configurations, with $\tm = \zd,\ \sd$ being perfect crystals, and $\tm =  32.2^\circ$ being the $\Sigma_{13}$ GB. The GBs vicinal to these high symmetry configurations have structures that are nominally the same as the high symmetry configuration, plus additional (or missing) dislocations separated by large distances ($\sim |\mathbf{b}|/\delta \tm$). These well separated `perturbations' produce out of plane distortions that need a very large cell to relax completely. Our simulations use a 1000\ \AA\ wide cell (in the direction perpendicular to the GB), and while we see significant decrease in energy over small cells (we have studied cells with widths of 50-1000\ \AA) due to relaxation, even the 1000\ \AA\ wide cell is not sufficiently large enough to fully relax the GB energy and obtain the infinite cell size limit. Note that this problem exists mostly for GBs that are vicinal to high symmetry configurations. 

Our model for GB energy is based on a `small perturbation' approach; however, all the evidence presented so far provides only indirect validation of the perturbation idea. The perturbation model can be supported by inspecting GBs in the vicinity of the high symmetry $\Sigma_{7,\ 13}$ boundaries. We consider both symmetric and asymmetric perturbations, as shown in Figure~\ref{fig:figure5}. This figure (plots a-d) shows that GBs in the vicinity of the $\Sigma_7$ GB have basically the same structure as $\Sigma_7$, plus an occasional extra (or missing) dislocation, as the case might be. The asymmetric perturbation ($\delta \tl \neq 0$) sometimes results in a faceted boundary (Figures~\ref{fig:figure5}d, h), with the facet locally following the high symmetry GB. The kinks joining the facets are composed of extra dislocations that are not present in the high symmetry GB. The same observations are true for the GBs vicinal to the $\Sigma_{13}$ GB. It is remarkable that the GB generation algorithm is able to capture all the features expected from well annealed graphene GBs.

We have found that all of the approximately 79,000 lowest-energy configuration GBs that we have simulated consist of only pentagon-heptagon pairs, and the usual hexagonal rings.  No other geometric configurations were observed for the lowest energy boundaries. For example, the 5-8-5 configurations that have been previously observed experimentally at grain boundaries \cite{Lahiri2010326, Yang201412041, Ma2014226802} were not found in our structures. From an energy point of view, the 5-8-5 defects are vacancy defects
and thus should be precluded from the ground state structures. However, non-equilibrium structures, such as the 5-8-5 defects, can indeed be captured by our algorithm if the Hamiltonian (Equation 2 in Ref.~\cite{ophus2015large}) is not driven to its minima (the convergence criteria could be suitably relaxed, or Metropolis sampling could be performed at a suitably defined ``temperature''). Further, the absence of such defects from our GBs is consistend with the fact that, to the best of our knowledge, such defects have not been observed in free-standing graphene films. Rather, they have been observed either in films on substrate or in free-standing graphene films after electron beam irradiation has modified their structure \cite{Kotakoski2011245420}.  The ubiquity of pentagon-heptagon pairs in graphene grain boundaries is consistent with our our HRTEM study of 176 boundaries~\cite{ophus2015large}, the majority of which did not contain any rings of more than 7 or less than 5 carbon atoms. Further, the GB generation algorithm is able to capture faceting where appropriate.



To conclude, the main contribution of this work is to develop a fundamental understanding of the structure and energy of the entire space of graphene GBs. We have developed analytical expressions for GB energy as functions of the misorientation and line angle that can be readily used in future calculations of grain growth or other GB related phenomena~\cite{Mishin20101117}. We hope that our analysis will pave the way for a deeper understanding of GB interfaces in graphene and other 2D materials. 

\section*{Acknowledgement}
A.S.~thanks R.~O.~Ritchie for hosting him at LBNL, and the Miller Institute for Basic Research in Science at University of California Berkeley for providing financial support via the Miller Research Fellowship. The work was supported at LBNL by the Mechanical Properties of Materials Program (KC-13) funded by the U.S. Department of Energy, Office of Science, Office of Basic Energy Sciences, Materials Sciences and Engineering Division under contract no. DE-AC02-05CH11231.  C.O.~acknowledges the Molecular Foundry, supported by the Office of Science, Office of Basic Energy Sciences, of the U.S.~Department of Energy under Contract no. DE-AC02-05CH11231.

\section*{Methods}
The GB structures used in this study were generated by using the algorithm introduced in Ref.~\cite{ophus2015large}, and are available online~\cite{onlineRepo}. In order to minimize boundary effects, we used GB structures that were 1000\ \AA\ wide, thus the GB was 500\ \AA\ from the boundaries. All GB structures used in this study are periodic in the direction parallel to the GB, and had a maximum length of 2000\ \AA. The atoms were allowed to relax in the out of plane direction without any constraint; a plot of the maximum out of plane displacement as a function of the misorientation and the line angle can be found in Supporting Information Section S4. The simulations were done in the LAMMPS~\cite{plimpton1995} code with the AIREBO interatomic potential~\cite{stuart2000,stuart2002}. Atoms in a strip of width 10\ \AA\ on the edges of the sample were held fixed at their ideal lattice positions during the simulations in order to reduce the boundary effects. Each sample was prepared by first relaxing the atoms with the conjugate gradient (CG) algorithm  so that the force on each atom was less than 0.01 eV/\AA. The sample was then held at 300 K for 10 picoseconds by simulating the NVT ensemble (fixed Number of atoms, Volume, and Temperature). This step was used to introduce any out of plane deformation that might have been missed by the CG algorithm. Finally, the atoms were again relaxed to within a residual force of 0.01 eV/\AA\ with the CG algorithm. The atoms at the strips on the edges were held fixed throughout these steps. As mentioned previously, the energy of the GB was measured as 
$\gamma(\tm,\tl) = (E_{\mathrm{total}} - n_{\mathrm{atoms}}E_{\mathrm{bulk}})/l_\mathrm{GB}$,
where $E_{\mathrm{total}}$ is the net energy of the configuration, $n_{\mathrm{atoms}}$ is the number of atoms in the configuration, $E_\mathrm{bulk}$ is the energy per-atom  in the reference crystal (= -7.81 eV for the AIREBO potential), and $l_\mathrm{GB}$ is the length of the GB. The atoms that were held fixed and not allowed to relax were not included in the energy calculations. The linear elastic constants $G$ and $\mu$ were calculated with the widely used technique of imposing small deformations on a relaxed bi-periodic graphene crystal, measuring the energy, and fitting the measured energy to the energy expression from linear elastic theory. This method yields $G$ = 325.68 GPa, and $\mu$ = 0.318. The fits of the measured GB energy to Equations~\ref{eq:symGBEnergy},~\ref{eq:FullGBEnergy} were done with a standard least-squares algorithm.

\bibliography{references} 
\bibliographystyle{plain} 

\end{document}